\def\year{2018}\relax
\documentclass[letterpaper]{article} 
\usepackage{aaai18}  
\usepackage{times}  
\usepackage{helvet}  
\usepackage{courier}  
\usepackage{url}  
\usepackage{graphicx}  

\usepackage[utf8]{inputenc} 
\usepackage[T1]{fontenc}    
\usepackage{url}            
\usepackage{booktabs}       
\usepackage{amsfonts}       
\usepackage{nicefrac}       
\usepackage{microtype}      
\usepackage{times}
\usepackage{amsmath}
\usepackage{amsthm}
\usepackage{amssymb}
\usepackage{subfig} 
\usepackage{bm} 
\usepackage{mdframed}
\usepackage{multirow}
\usepackage{tikz}
\usepackage{mathtools}
\usepackage{caption}
\usepackage{natbib}

\usepackage[colorlinks=true,bookmarks=true, pdftex, citecolor = black, linkcolor = black, urlcolor = black]{hyperref}
\hypersetup{%
	breaklinks,
	baseurl       = http://,
	pdfborder     = 0 0 0,
	pdfpagemode   = UseNone,
	pdfpagelabels = false,
	pdfstartpage  = 1,
	bookmarksopen = true,
	bookmarksdepth= 3,
}

\DeclareMathOperator*{\argmax}{arg\,max}

\newcommand{\ds}{\displaystyle}

\newcommand{\F}{\mathcal{F}}

\newcommand{\X}{\mathcal{X}}
\newcommand{\mN}{\mathcal{N}}
\newcommand{\mL}{\mathcal{L}}

\newcommand{\Reals}{\mathbb{R}}
\newcommand{\cond}{\:|\:}

\newcommand{\pos}[1]{\left( #1 \right)_+}
\newcommand{\bw}{\bm{w}}
\newcommand{\bx}{\bm{x}}
\newcommand{\bc}{\bm{c}}
\newcommand{\bb}{\bm{b}}
\newcommand{\by}{\bm{y}}

\newcommand{\hP}{\hat{P}}

\newcommand{\paths}{\texttt{paths}}
\newcommand{\regions}{\texttt{regions}}
\newcommand{\arbitrary}{\texttt{arbitrary}}
\newcommand{\scheduling}{\texttt{scheduling}}

\newtheorem{proposition}{Proposition}
\newtheorem{lemma}{Lemma}
\newtheorem{theorem}{Theorem}

\begin{document}
	%
	\title{A Bayesian Clearing Mechanism for Combinatorial Auctions}
	\author{
		Gianluca Brero\\
		University of Zurich\\
		{\small\tt brero@ifi.uzh.ch}
		\And
		S\'{e}bastien Lahaie\\
		Google Research\\
		{\small\tt slahaie@google.com}
	}
	\maketitle
	
	\begin{abstract}
		We cast the problem of combinatorial auction design in a Bayesian
		framework in order to incorporate prior information into the
		auction process and minimize the number of rounds to convergence.
		We first develop a generative model of agent valuations and market
		prices such that clearing prices become maximum a posteriori
		estimates given observed agent valuations. This generative model
		then forms the basis of an auction process which alternates
		between refining estimates of agent valuations and computing
		candidate clearing prices. We provide an implementation of the
		auction using assumed density filtering to estimate valuations and
		expectation maximization to compute prices. An empirical
		evaluation over a range of valuation domains demonstrates that our
		Bayesian auction mechanism is highly competitive against the
		combinatorial clock auction in terms of rounds to convergence,
		even under the most favorable choices of price increment for this
		baseline.
	\end{abstract}
	
	
	\section{Introduction}
	
	Combinatorial auctions address the problem of allocating multiple
	distinct items among agents who may view the items as complements or
	substitutes. In such auctions, agents can place bids on entire
	packages of items in order to express complex preferences, leading to
	higher allocative efficiency. Nevertheless, bidding in a combinatorial
	auction places a substantial cognitive burden on agents, because the
	process of valuing even a single bundle can be a costly
	exercise~\citep{kwasnica2005new,parkes2006mit}. There is therefore
	great interest in developing \emph{iterative} combinatorial auctions,
	which help to guide the bidding process using price feedback, and in
	devising techniques to limit the number of rounds needed to reach
	convergence (ideally in the dozens rather than
	hundreds)~\citep{petrakis2012ascending,bichler2017coalition}.
	
	In this work, we propose to incorporate \emph{prior information} on
	agent valuations into the auction procedure in a principled manner,
	thereby achieving a low number of rounds in practice. We cast the
	problem of combinatorial auction design in a Bayesian framework by
	developing a joint generative model of agent valuations and market
	prices. 
	Our generative model defines a likelihood function for clearing prices
	given agent valuations. If these valuations are observed, the maximum 
	a posteriori (MAP) estimate for prices corresponds to market clearing prices. 
	If they remain latent, valuations can be marginalized away,
	weighed by their own likelihood according to observed bids. This
	forms the basis for an auction scheme to solve the more general
	clearing problem where valuations are unknown.
	
	We consider settings where several indivisible items are up
	for sale, and agents have super-additive valuation functions
	over bundles of items (i.e., the items are pure complements). We
	provide an auction implementation using item prices consisting of
	two components. In the \emph{knowledge update} component, we
	maintain a Gaussian posterior over agent valuations, which is
	updated as new bids are placed using assumed density
	filtering~\citep{opper1998bayesian}. Prior information can be
	incorporated into the auction by suitably initializing this
	component.
	The knowledge update step presumes that agents follow \emph{myopic
		best-response} strategies and bid on utility-maximizing bundles at
	each round. Accordingly, we discuss an extension to our auction
	scheme using multiple price trajectories that incentivizes this
	behavior in \emph{ex post} Nash equilibrium.
	In the \emph{price update} component, we obtain an
	analytical expression for the clearing price objective, based on the
	Gaussian model of valuations that the auction maintains. We
	establish that the form of the objective is suitable for
	optimization using expectation maximization. By alternating the two
	components, we obtain an intuitive and tractable auction scheme
	where agents place bids, knowledge over latent valuations is updated
	given bids, and prices are updated given current knowledge of
	valuations.
	
	For evaluation purposes, we first illustrate our auction on a
	stylized instance to gain insight into the auction's behavior under
	both unbiased and biased prior information. We then conduct
	simulation experiments to compare our auction implementation against
	a combinatorial clock auction that updates prices according to
	excess demand, which is the standard price update scheme used in
	practice~\citep{ausubel2014practical}. The prior information in our
	Bayesian auction is obtained by fitting a Gaussian process prior on
	a training sample of valuations.
	The baseline clock auction is parametrized by a
	step size, or price increment. We find in our experiments that our
	Bayesian auction is competitive against the strongest possible version
	of the baseline auction, where the price increment is chosen
	separately for each instance to lead to the fewest possible rounds.
	In particular, the Bayesian auction almost matches the strongest possible version
	of baseline auction in terms of
	number of instances cleared, and uses fewer rounds on average when it
	is able to clear.

	\section{Preliminaries}
	\label{sec:preliminaries}
	
	We consider a setting with $m$ distinct and indivisible items, held by
	a single seller. The items are to be allocated among $n$ agents (i.e.,
	buyers). We will use the notation $[n] = \{1,\dots,n\}$, so that $[n]$
	and $[m]$ denote the index sets of agents and items, respectively.
	There is unit supply of each item. A \textit{bundle} is a subset of
	the set of items. We associate each bundle with its indicator vector,
	and denote the set of bundles as $\X=\{0,1\}^m$. The component-wise
	inequality $x \leq x'$ therefore means that bundle $x$ is contained in
	bundle $x'$. The empty bundle is denoted by $\emptyset$.
	
	Each agent $i$ is \emph{single-minded} 
	so that its valuation can be
	encoded via a pair $(x_i, w_i)$ where $x_i \in \X$ is a bundle and
	$w_i \in \Reals_+$ is a non-negative value (i.e., willingness
	to pay) for the bundle.
	The agent's valuation function $v_i: \X \rightarrow \Reals_+$ is defined as
	$v_i(x)=w_i$ if $x \geq x_i$, and $v_i(x)=0$ otherwise.
	%
	%
	In words, the agent only derives positive value if it acquires
	all the items in $x_i$ (which are therefore complements), and
	any further item is superfluous.
	Our auction and results all extend to agents with OR
	valuations, which are concise representations of super-additive
	valuations~\citep{nisan2000bidding}.\footnote{More formally, an OR
		valuation takes the form
		$v(x) = \max\{v_1(x_1) + v_2(x_2) : x_1+x_2=x,\: x_1 \wedge x_2
		= 0 \}$, where $v_1$ and $v_2$ are themselves OR valuations or
		single-minded.} This is due to the fact that an agent with an OR
	valuation will behave and bid in our auction exactly like a set of
	single-minded agents, under myopic
	best-response~\citep{parkes1999bundle}. Under super-additive
	valuations, items are pure complements, and complementarities are
	a key motivation for using package bidding. For the sake of
	simplicity, however, we limit the exposition to single-minded
	agents.
	
	An \emph{allocation} is represented as a vector of bundles
	$\by = (y_1,\dots,y_n)$, listing the bundle that each agent obtains
	(possibly $\emptyset$).
	An allocation is \emph{feasible} if the listed bundles are pairwise
	disjoint (i.e., each item is allocated to at most one agent). We
	denote the set of feasible allocations by $\F$.
	%
	The purpose of running a combinatorial auction is to find an
	\emph{efficient} allocation of the items to the agents, meaning an
	allocation that maximizes the total value to the
	agents.\footnote{This is in contrast to the goal of maximizing
		\emph{revenue}. In auction design, one typically begins with an
		efficient auction, which is then modified (e.g., using reserve
		prices) to achieve optimal revenue~\citep{myerson1981optimal}. We
		therefore consider the problem of designing an efficient auction
		as more fundamental.}
	More formally, a feasible allocation
	$\by \in \F$ is efficient if
	$
	\by \in \argmax_{\by' \in \F} \sum_{i \in [n]} v_i(y'_i).
	$
	However, an iterative auction proceeds via a price adjustment process,
	so prices will be our central object of study, rather than
	allocations. The allocation in an iterative auction is adjusted
	according to agents' responses to prices.

	\paragraph{Clearing Prices}
	In the context of a combinatorial auction, we encode prices as a
	non-negative function $\theta: \X \rightarrow \Reals_+$ over the
	bundles. We assume that prices are normalized and monotone:
	$\theta(\emptyset) = 0$, and $\theta(x) \le \theta(x')$ if $x\le x'$.
	An iterative auction adjusts prices to
	balance demand and supply. To formalize this notion, we need several
	additional concepts.
	We assume that agents have quasi-linear utility, so that the utility
	to agent $i$ of obtaining bundle $x$ at prices $\theta$ is
	$v_i(x) - \theta(x)$. The \emph{indirect utility} function provides
	the maximum utility that agent $i$ can achieve, when faced with prices
	$\theta$, by choosing among bundles from $\X$:
	\begin{equation} \label{eq:indirect-buyer}
	V_i(\theta) = \max \left\{\, v_i(x) - \theta(x) : x \in \X \,\right\}.
	\end{equation}
	Note that for single-minded agents, the indirect utility reduces to
	$V_i(\theta; x_i, w_i) = \pos{w_i - \theta(x_i)}$, where the notation
	$\pos{a} = \max\{a, 0\}$ refers to the positive part of the argument.
	It will sometimes be useful to make explicit the parametrization of
	the indirect utility on the agent's type $(x_i, w_i)$, as we have just
	done. The \emph{demand set} of agent $i$ is defined as
	$D_i(\theta) = \left\{x \in \X : v_i(x) - \theta(x) = V_i(\theta)
	\right\}$.
	Similarly, the \emph{indirect revenue} function provides the maximum
	revenue that the seller can achieve, when faced with prices $\theta$,
	by selecting among feasible allocations:
	\begin{equation} \label{eq:indirect-seller}
	R(\theta) = \max \left\{\, \sum_{i \in [n]} \theta(y_i) : \by \in \F \,\right\}.
	\end{equation}
	The seller's \emph{supply set} consists of the feasible allocations
	that maximize revenue:
	$S(\theta) = \{ \by \in \F : \sum_{i \in [n]} \theta(y_i) = R(\theta) \}.$
	
	We say that prices $\theta$ are \emph{clearing prices} if there is a
	feasible allocation $\by$ such that, at prices $\theta$, the seller's
	revenue is maximized, and each agent's utility is maximized. Formally,
	we require the following conditions: $\by \in S(\theta)$ and
	$y_i \in D_i(\theta)$ for all $i \in [n]$. We say that the clearing
	prices $\theta$ $\emph{support}$ allocation $\by$.
	
	
	It is a standard result that the set of allocations supported by any
	given clearing prices $\theta$ coincides with the set of efficient
	allocations. (This is a special case of the Fundamental Theorems of
	Welfare Economics~\citep[16.C--D]{mas1995microeconomic}.)
	Moreover,
	\citet{bikhchandani2002package} have shown that clearing prices exist
	and coincide with the minimizers of the following objective function, 
	which corresponds to the linear programming dual of the problem of allocating the items efficiently:
	\begin{equation}\label{eq:clearing-pot}
	\sum_{i \in [n]} V_i(\theta) + R(\theta) .
	\end{equation}
	This is a piece-wise linear, convex function of the price function
	$\theta$. Importantly, this result is guaranteed to hold only if the
	prices are an unrestricted function over the bundles (except for
	non-negativity and normalization). In practice, it is common to use
	certain parametrizations for the prices. For instance, taking
	$\theta(x) = p \cdot x$ for some vector $p \in \Reals^m_+$ corresponds
	to using linear prices (i.e., item prices). These parametrizations may
	not achieve the unrestricted minimum in~\eqref{eq:clearing-pot}; in
	particular, linear clearing prices may not exist. We will use
	unrestricted prices in the development of our auction, and postpone
	the question of price parametrization until needed to achieve a
	practical implementation.
	
	It is useful to view~\eqref{eq:clearing-pot} as a potential function
	that quantifies how close prices $\theta$ are to supporting an
	efficient allocation. Indeed, if some prices achieve a value
	of~\eqref{eq:clearing-pot} that differs from the optimum by an
	additive error of $\delta$, then the agents (and seller) can be
	induced to accept an efficient trade using transfers totaling
	$\delta$.
	In the following, we will therefore refer to the function
	\begin{equation}\label{eq:clearing-potential}
	U(\theta; \bm v) = \exp{\left[-\sum_{i \in [n]} V_i(\theta) - R(\theta)\right]} 
	\end{equation}
	as the \textit{clearing potential} for the valuation profile
	$ \bm v = (v_1,...,v_n)$, which will capture, in a formal sense, how
	likely a price function $\theta$ is to clearing valuation profile
	$\bm v$.
	\paragraph{Iterative Auction and Incentives}
	
\begin{figure}[t!]
		\begin{center}
			\includegraphics[width=0.45\textwidth]{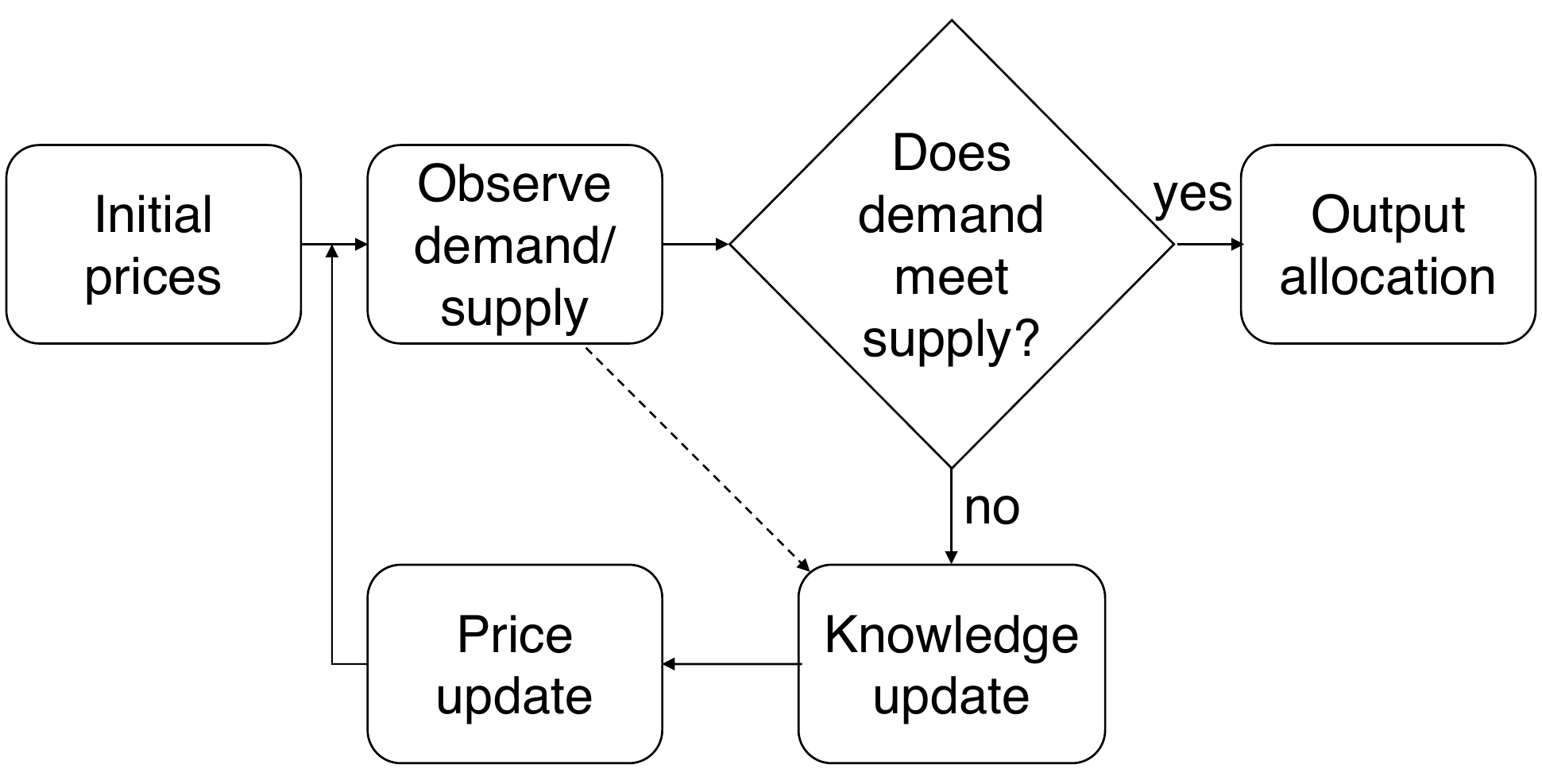}
		\end{center}
		\caption{Bayesian iterative auction.}
		\label{fig:BayesianClearingMechanism}
	\end{figure}

	The goal of our paper is to design an iterative auction that
	exploits the auctioneer's prior knowledge over agent valuations in
	order to speed up the clearing process. The auction proceeds over
	rounds. Agents report their demand at the current prices and, if
	the market is not cleared, the information provided by the agents
	is  used to update the knowledge about their valuations.
	Candidate clearing prices are computed based on the updated
	knowledge, and the procedure iterates. A schematic representation
	of the auction process is presented in
	Figure~\ref{fig:BayesianClearingMechanism}. The knowledge update
	and price update components constitute the core of the auction
	that must be implemented.
	
	The correctness of our auction relies on the agents following a
	strategy of \emph{myopic best-response} bidding, meaning that each
	agent bids on a utility-maximizing bundle at each round.
	%
	There is evidence that myopic bidding may be a reasonable assumption
	in practice. For instance, in the FCC broadband spectrum auction,
	jump bids were the exception~\citep{cramton1997fcc}. Nonetheless, a
	robust auction design should incentivize agents to follow the
	appropriate strategies. For this purpose, we can use an extension of
	our auction that maintains $n+1$ price trajectories in order to
	compute clearing prices when all agents are present, and when each
	agent is removed in turn. This allows one to compute final VCG
	payments and bring myopic best-response bidding into an \emph{ex post} Nash
	equilibrium~\citep{gul2000english,bikhchandani2006ascending}. The
	technique of using multiple trajectories was previously used
	by~\citet{ausubel2006efficient} and~\citet{mishra2007ascending}
	among others. We will provide a more precise treatment of incentives
	in the formal description of our auction mechanism.

	\section{Generative Model}
	\label{sec:generative-model}
	
	The purpose of this section is to define a probabilistic
	relationship between prices and valuations that will allow us
	to use the auctioneer's prior knowledge over valuations to
	make inferences over clearing prices. We write
	$\bw = (w_1,\dots,w_n)$ and $\bx = (x_1,\dots,x_n)$ for the
	vectors of agents' values and bundles, and denote the
	probabilistic model as $P(\bx,\bw,\theta)$. Below, our
	convention is that $Q$ refers to distributions---possibly
	unnormalized---that form the building blocks of the generative
	model, whereas $P$ refers to the normalized distribution
	resulting from the generative model. We represent the prior
	knowledge of the auctioneer over agent valuations via the
	probability density function
	$Q(\bm w) = \prod_{i\in[n]}Q(w_i)$.
	
	The structure of our probability model is inspired by the work
	of~\citet{sollich2002bayesian}, who provides a Bayesian
	interpretation of the support vector machine (SVM) objective. 
	To establish a proper relationship between prices and
	valuations, the key is to require that
	\begin{equation}\label{eq:generative-model}
	P(\bx, \bw, \theta) \propto U(\theta; \bx, \bw)Q(\bw),
	\end{equation}
	where $U(\theta; \bx, \bw)$ is the clearing potential
	introduced in \eqref{eq:clearing-potential}, adapted to
	single-minded valuations.
	Under this joint probability model, we have that the posterior probability of prices
	$\theta$ takes the form
	\begin{equation} \label{eq:posterior}
	P(\theta \cond \bx, \bw) \propto U(\theta; \bx, \bw).
	\end{equation}
	Therefore, the maximum a posteriori (MAP) estimate maximizes
	$U(\theta; \bx, \bw)$, or equivalently
	minimizes~(\ref{eq:clearing-pot}), and corresponds to clearing prices.
	
	To establish that a probability model of the
	form~(\ref{eq:generative-model}) is possible---namely, that it can
	indeed be normalized---we will derive it as the result of a
	generative model. This process may be of independent interest
	as a means of generating agents together with market prices.
	%
	%
	\begin{enumerate}
		\item Draw prices $\theta$ according to
		$Q(\theta) \propto \exp\left[ -R(\theta) \right].$
		\item For each agent $i \in [n]$:
		\begin{itemize}
			\item[--] Draw $w_i \in \Reals_+$ from $Q(w_i)$.
			\item[--] Draw $x_i \in \X$ from
			$$ Q(x_i \cond w_i, \theta) = \frac{1}{2^m}
			\exp \left [ -V_i(\theta; x_i, w_i) \right].$$
			\item[--] With probability $1 - \nu(w_i, \theta)$, restart from step 1, where $$ \nu(w, \theta) = \sum_{x \in \X} Q(x \cond w, \theta).$$			
		\end{itemize}
	\end{enumerate}
	Above, we must ensure that the $Q(\theta)$ prior normalizes; this is
	the case under our assumption that the domain of $\theta$ falls within
	the positive orthant.
	The prior distribution $Q(w_i)$ on value $w_i$ is
	left free in the model, so that it may correspond to the
	auctioneer's prior in practice.
	Note that the bundle
	likelihood $Q(x_i \cond w_i, \theta)$ is \emph{not} normalized;
	because $V_i(\theta; x_i, w_i) \geq 0$, summing over the set of
	bundles leads to the aggregate probability mass
	$\nu(w_i, \theta) \leq 1$. Rather than normalizing by this quantity,
	we use the ``remaining probability'' $1 - \nu(w_i, \theta)$ of not
	drawing any bundle to restart the process. Because of the possible
	restart, the agent types (bundle-value pairs) and clearing prices are
	not independent in the overall generative distribution. In particular,
	the number of agents $n$ in the economy affects the distribution of
	prices.

	The following proposition confirms that our model satisfies \eqref{eq:generative-model}.
	All proofs are deferred to the appendix.
	
	%
	\begin{proposition}\label{prop:gen-model}
		The generative model of agent types and prices takes the form
		\begin{equation}\label{eq:gen-model}
		P(\bx, \bw, \theta) \propto U(\theta; \bx, \bw)Q(\bw).
		\end{equation}
	\end{proposition}
	%

	\noindent
	The generative process defines a probability distribution over
	prices once valuations are \emph{observed}, but during the auction
	the valuations remain latent, and must be inferred based on
	observed bids placed across rounds. Under appropriate
	incentives, the auctioneer can infer valuations assuming that
	the agents follow myopic best-response bidding.
	However, if there are any bidding errors or corruption in
	communication, assuming exact best-response can cause
	singularities in the inference process (e.g., there may be no
	valuation consistent with all observed bids). To guard against
	this, our mechanism will integrate bids as if they were 
	generated from the following stochastic model: Let $b_i \in \{-1,+1\}$ be an
	indicator variable to denote whether the agent bids on bundle
	$x_i$ ($b_i = +1$) or not ($b_i = -1$); the latter is equivalent
	to bidding on $\emptyset$. If the cost of bundle $x_i$ is $c_i$,
	then the choice of bid follows the probability distribution
	\begin{equation}\label{eq:bid-behavior}
	Q(b_i = +1 \cond  c_i, w_i) \propto \Phi(\beta(w_i - c_i)),
	\end{equation}
	where $\Phi$ is the cumulative distribution function of the
	standard normal, and $\beta > 0$ is a scalar parameter. This
	is known as the \emph{probit} variant of approximate
	best-response, which arises from random utility
	models~\citep{train2009discrete}.
	As $\beta \rightarrow \infty$, we obtain exact best-response: the
	agent bids on $x_i$ if and only if
	this bundle yields positive utility under bundle cost $c_i$.
	Using a large (but finite) $\beta$ allows the auctioneer to
	model agents as essentially following a best-response
	strategy, but occasionally allowing for bidding errors or inconsistencies.
	
	
	\section{Auction Description}
	\label{sec:AuctionDescription}
	
	Our auction proceeds over rounds; we use $k$ to denote the current
	round, and $\ell$ to index the rounds up to $k$. At each round, prices
	are updated, which imputes a cost to each agent's bundle. Let
	$c_i^\ell$ be the cost of agent $i$'s desired bundle $x_i$ in round
	$\ell$ according to the current prices. The prices at each round
	should not be confused with the latent clearing prices $\theta$, which
	we are trying to compute as a MAP estimate of the generative model.
	Given its value $w_i$ and the bundle cost $c_i^\ell$, agent $i$ places
	bid $b_i^\ell \in \{-1,+1\}$ in round $\ell$.
	
	We write $\bc^\ell = (c_1^\ell, \dots, c_n^\ell)$ to denote the vector
	of bundle costs in round $\ell$, and
	$\bc^{(\ell)} = (\bc^1, \dots, \bc^\ell)$ to denote the vector of
	costs \emph{up to} round $\ell$. For brevity we also write $\bc = \bc^{(k)}$
	to denote the vector of all costs up to the current round. We use the
	notation $\bb^\ell$, $\bb^{(\ell)}$, and $\bb$ to denote the analogous
	vectors of bids.
	The bundle costs and agent bids in a round depend on the
	current prices, which themselves depend on the bids placed in all
	earlier rounds. Assuming that the first round prices are zero, we have
	the following intuitive posterior over bids and costs.

	\begin{lemma}\label{lem:bids-cost-posterior}
		The posterior distribution over bids and costs placed during the
		auction, given the generated prices and agent types, is given by
		\begin{equation*}
		P(\bb, \bc \cond \bx, \bw, \theta) = Q(\bb \cond \bc, \bw)
		\prod_{\ell=1}^k P(\bc^\ell \cond \bb^{(\ell-1)}, \bc^{(\ell-1)}, \bx),
		\end{equation*}
		where $\bb^\ell$ and $\bc^\ell$ are the vectors of agent bids and
		costs at round $\ell$, and $\bb^{(\ell)}$ and $\bc^{(\ell)}$ are the
		vectors of agent bids and costs \emph{up to} round $\ell$.
	\end{lemma}
	We see that the posterior over bids and costs does not depend
	on the underlying clearing prices $\theta$, conditional on
	agent types $(\bx, \bw)$, because the initial prices and agent
	valuations fully determine how the auction proceeds. More
	specifically, the posterior decomposes into the likelihood of
	the observed bids $Q(\bb \cond \bc, \bw)$ under stochastic
	model~\eqref{eq:bid-behavior}, times the likelihood of the
	observed sequence of costs. The latter does not involve $\bw$,
	because current round prices are fully determined by the bids
	and costs of previous rounds.
	Our auction is based on the following characterization of
	the overall posterior over prices and agent values.
	\begin{proposition}\label{prop:full-posterior}
		The posterior distribution of latent variables $(\bw, \theta)$ given
		observed variables $(\bb, \bc, \bx)$ takes the form
		$$
		P(\bw, \theta \cond \bb, \bc, \bx) \propto Q(\bb \cond \bc, \bw)Q(\bw)
		\cdot Q(\bx \cond \bw, \theta) Q(\theta),
		$$
		where the proportionality constant depends solely on $(\bb, \bc,
		\bx)$.
	\end{proposition}
	The posterior factors into two terms, which motivates our auction
	procedure. The first term, $Q(\bb \cond \bc, \bw)Q(\bw)$, can be
	construed as a posterior over agent values given bids and costs, since
	$Q(\bw)$ corresponds to a prior and $Q(\bb \cond \bc, \bw)$
	corresponds to a likelihood. 
	We will
	maintain an approximation $\hP(\bw)$ to this posterior over agent
	values and update it as new bids are placed in response to bundle
	costs. This is the \emph{knowledge update} component.
	
	Recalling~\eqref{eq:posterior}, the second term
	$Q(\bx \cond \bw, \theta) Q(\theta) \propto U(\theta; \bx, \bw)$ in the posterior corresponds (up
	to a constant factor) to the price posterior given knowledge of agent
	types. This leads to an approximation to the price posterior when
	values remain latent:
	\begin{equation}\label{eq:price-map}
	P(\theta \cond \bb, \bc, \bx) \approx \int \mathrm{d}\bw\, \hP(\bw)
	\cdot Q(\bx \cond \bw, \theta) Q(\theta).
	\end{equation}
	Here we have simply integrated the full posterior as given by
	Proposition~\ref{prop:full-posterior}, and made use of our
	approximation to the value posterior. (We have also omitted the
	normalization constant.) In the context of an auction, we quote a
	specific price function to the agents, rather than a distribution over
	prices. Therefore, in the \emph{price update} component, we will
	compute and quote the MAP estimate of prices by maximizing~\eqref{eq:price-map}.
	Note that if we have exact knowledge of agent values (i.e., $\hP$ is a
	point mass), computing the MAP estimate is equivalent to
	minimizing~(\ref{eq:clearing-pot}) and to computing clearing
	prices, as one would expect.

	\paragraph{Knowledge Update} We observe that the value posterior
	$Q(\bb \cond \bc, \bw)Q(\bw)$ consists of a separate factor for each
	agent~$i$, taking the form
	$
	Q(w_i) \prod_{\ell=1}^k Q(b_i^\ell \cond c_i^\ell, w_i),
	$
	where $k$ is the current round. This represents a posterior on agent
	$i$'s individual value $w_i$. To obtain an approximation to this
	posterior, we use an online scheme known as \emph{assumed density
		filtering}, which is a special case of expectation
	propagation~\citep{cowell1996comparison,minka2001family,opper1998bayesian}.
	Under this approach, a Gaussian distribution
	$\hP(w_i; m_i, \sigma^2_i)$ is used to approximate the posterior; its
	mean $m_i$ and variance $\sigma^2_i$ are updated at each round given
	the bidding observations. The Gaussian is initially set to approximate
	the prior $Q(w_i)$ via moment matching: $m_i$ and $\sigma^2_i$ are set
	to the mean and variance of this prior $Q$. In each later round
	$\ell = 1,\dots,k$ the posterior is again updated by matching the
	moments of
	$
	Q(b_i^\ell \cond c_i^\ell, w_i) \hP(w_i; m_i, \sigma^2_i),
	$
	which is an online update.
	Using moment matching as an approximation is justified by the fact
	that it corresponds to minimizing the Kullback-Leibler divergence
	$\mathrm{KL}(Q \| \hP)$ under the constraint that $\hP$ is Gaussian.\\	
	Due to the form of the likelihood~\eqref{eq:bid-behavior} and the fact
	that $\hP$ is Gaussian, the update has a closed-form
	solution~\citep[see ][p.~74]{williams2006gaussian}:
	\begin{eqnarray*}
		m_i & \leftarrow & m_i + \frac{b_i^\ell\, \sigma_i^2\, \beta
			\mN(z_i)}{\sqrt{1+\sigma_i^2\beta^2} \, \Phi(z_i) } \\
		\sigma^2_i & \leftarrow & \sigma^2_i - \frac{\sigma_i^4\, \beta^2 \mN(z_i)}{1+\sigma_i^2\beta^2 \Phi(z_i) }\left(z_i+\frac{\mN(z_i)}{\Phi(z_i)}\right),
	\end{eqnarray*}
	where $\mN$ and $\Phi$ are the probability density and cumulative
	distribution functions of the standard normal, respectively, and where
	$z_i = b_i^\ell \,\beta(m_i - c_i^\ell) / \sqrt{1+\sigma_i^2\,\beta^2}$.
	Recall that $\beta$ is a positive parameter characterizing the 
	extent to which the auctioneer assumes that agents make
	mistakes in placing best-response bids.
	Since $\beta$ is positive, the mean $m_i$ is updated in the
	direction of the bid $b_i^{\ell}$.
	On the other hand, the variance $\sigma^2_i$ is strictly
	decreasing, thus ensuring that the beliefs over bidder values
	converge to a point mass in the limit as the rounds progress,
	and that the auction converges to a final vector of prices.

	
	\paragraph{Price Update}
	
	To implement the price update component we need an algorithm to
	maximize the approximate posterior~\eqref{eq:price-map}. This
	posterior factors into $Q(\theta)$ and a term for each agent $i$, which
	we denote as
	$$
	\mL_i(\theta) = \int \mathrm{d}w_i\, \hP_i(w_i; m_i, \sigma_i^2) Q(x_i \cond w_i, \theta).
	$$
	Because $Q_i(x_i \cond w_i, \theta)$ has an exponential form, and
	$\hP_i$ is a Gaussian, this integral has a closed form solution  (see appendix). 
	Let $q_i \in \{0,1\}$ be a binary auxiliary variable. We have
	$
	\mL_i(\theta)
	= \sum_{q_i = 0}^1 \mL_i(\theta, q_i),
	$
	where we define
	\begin{eqnarray*}
		\mL_i(\theta, 1) &\!\!\!\!\! = & \!\!\!\!\! \Phi\left( \frac{m_i - \theta(x_i)}{\sigma_i} - \sigma_i \right) \exp\left[ \theta(x_i) - m_i + \frac{\sigma_i^2}{2} \right] ,\\
		\mL_i(\theta, 0) &\!\!\!\!\! = & \!\!\!\!\! \Phi\left( \frac{\theta(x_i) - m_i}{\sigma_i} \right).
	\end{eqnarray*}
	Here $\Phi$ is again the cumulative distribution function of the
	standard normal. To summarize, taking the log of~\eqref{eq:price-map},
	the objective we seek to maximize with respect to $\theta$ is
	\begin{equation}\label{eq:price-objective}
	\log Q(\theta) + \sum_{i \in [n]} \log \sum_{q_i = 0}^1 \mL_i(\theta, q_i).
	\end{equation}
	Now, because $\Phi$ is log-concave, both $\mL_i(\theta, 0)$ and
	$\mL_i(\theta, 1)$ are log-concave in $\theta$. Ignoring the first
	term for an instant, we see that the objective consists of a sum of
	mixtures $\mL_i(\theta)$ of log-concave functions for each agent.
	This kind of objective is well-suited to optimization using the
	expectation-maximization (EM) algorithm~\citep{dempster1977maximum}.
	The $q_i$ amount to ``latent'' variables and the ``marginal''
	likelihood appears within the objective~\eqref{eq:price-objective}.
	(However, we do not claim any intuitive interpretation for the latent
	$q_i$---they are simply used to fit the objective into the mold of EM.)
	
	The remaining term is $\log Q(\theta)$, which is $-R(\theta)$ up to an
	additive constant. Recalling the definition of the seller's indirect
	utility~\eqref{eq:indirect-seller}, we see that this term is very
	complex for unrestricted $\theta$, because the set of feasible
	allocations $\F$ has a very complicated structure. To address this we
	will impose a linear structure on prices: $\theta(x) = p \cdot x$
	where $p \in \Reals^m_+$ denote item prices. With this
	parametrization, we have $R(\theta) = \sum_{j=1}^m p_j$, because any
	allocation that allocates all the items maximizes revenue under linear
	prices. The $\log Q(\theta)$ term therefore becomes a linear term in
	$p$, which is straightforward to incorporate within the EM algorithm.
	
	
	\paragraph{Incentive Compatibility}
	
	Our auction converges to an efficient allocation and
	clearing prices under myopic best-response bidding, but to
	ensure that agents follow such a strategy, they must be
	incentivized to do so. The standard technique used to achieve
	this in the literature on iterative auctions is to charge VCG
	payments upon
	completion~\citep{gul2000english,bikhchandani2006ascending}.
	But whereas VCG payments (together with an efficient
	allocation) induce truthful bidding in dominant strategies for
	single-shot auctions, weaker results hold for iterative auctions.
	
	A strategy profile constitutes an \emph{ex post} Nash
	equilibrium if no agent would like to deviate from its
	strategy, holding the others' strategies fixed, even with
	knowledge of the private valuations of the other agents.
	\citet{gul2000english} prove the following result:
	\begin{theorem}[\citeauthor{gul2000english}, \citeyear{gul2000english}]
		Truthful myopic best-response bidding is an \emph{ex post}
		Nash equilibrium in an iterative auction that myopically-implements
		the VCG outcome.
	\end{theorem}
	\noindent
	Above, the VCG outcome refers to an efficient allocation along
	with VCG payments, and an auction myopically-implements this
	outcome if the auction converges to it under myopic
	best-response bidding. The reason that truthfulness only holds
	in \emph{ex post} Nash equilibrium, rather than dominant
	strategies, is that profitable deviations may exist if another
	agent bids in a manner inconsistent with any valuation.
	
	Our auction already computes the efficient allocation under
	these conditions by virtue of converging to clearing prices.
	To compute VCG payments, we can simply extend or auction
	drawing on the idea of multiple
	price trajectories: the usual trajectory traced by our
	auction, and the trajectories that would result if each agent
	were removed in turn. This technique was previously used
	by~\citet{ausubel2006efficient}
	and~\citet{mishra2007ascending}.
	In this extended design, at each round, agents place bids against $n+1$
	different price vectors. Upon completion, the agents place
	last-and-final bids for their allocated bundles, thereby
	communicating their value for the allocations; importantly,
	agents do not need to communicate values for any bundles they
	did not win. This information is precisely what is needed to
	compute VCG payments~\citep[see, e.g., ][]{parkes2000preventing}.
	
	
	\section{Empirical Evaluation}
	\label{sec:empirical-evaluation}
	
	In this section we evaluate our Bayesian auction design with two
	different kinds of experiments: a small experiment to illustrate the
	behavior of our auction under biased and unbiased prior information,
	and a larger-scale experiment to compare our auction against a
	competitive baseline.
	
	Our simulations are conducted in Matlab. In all our experiments, we assume
	that agents best respond to the proposed prices (i.e., they always bid on their most 
	profitable bundle), and that the auctioneer considers their bids as if they were generated 
	from the response model presented in~(\ref{eq:bid-behavior}) with $\beta = 10$. 
	However, simulations where real bids follows~(\ref{eq:bid-behavior})
	with $\beta = 4$ provide results similar to the ones presented. For
	the price update, the objective~(\ref{eq:price-objective}) is
	maximized using the ``active-set'' algorithm in Matlab. To avoid
	numerical singularities we place a lower bound of 0.01 on the
	variance of valuation estimates.

	\paragraph{LLG Experiments} We consider an instance of the
	\emph{Local-Local-Global} (LLG) domain~\citep{ausubel2006lovely}, which has been considered
	several times in the combinatorial auctions
	literature.
	There are two items and three single-minded agents. Two of the
	agents are \emph{local}, meaning that they are interested in just one
	item, respectively the first and second item. The last agent is
	\emph{global} in the sense that it is interested in both items.
	
	The two local agents have a value of 4 for their respective items,
	and the global one has a value of 10 for both. The items are
	efficiently allocated when they are both assigned to the global
	agent, and linear prices are expressive enough to clear the market
	(e.g., we can use a price of 4 for each item).
	
	We assume that the auctioneer has accurate knowledge
	of the local agents' values: 
	$Q(w_1) = \mathcal N(w_1; m_1=4, \sigma_1^2=0.01)$ and 
	$Q(w_2) = \mathcal N(w_2; m_2=4, \sigma_2^2=0.01)$.
	Note the very low variance, reflecting certainty. We test how
	different kinds of prior knowledge over the global agent's
	value affect the number of rounds that the Bayesian auction takes to
	clear the market. In the first case we assume unbiased prior
	knowledge: $m_3=10$. In the second case we assume that it is biased
	below: $m_3=4$. Here, the auctioneer tends to
	allocate to the local agents instead of the global one.

\begin{figure}[t!]		
		\begin{center}
			\includegraphics[width=0.4\textwidth]{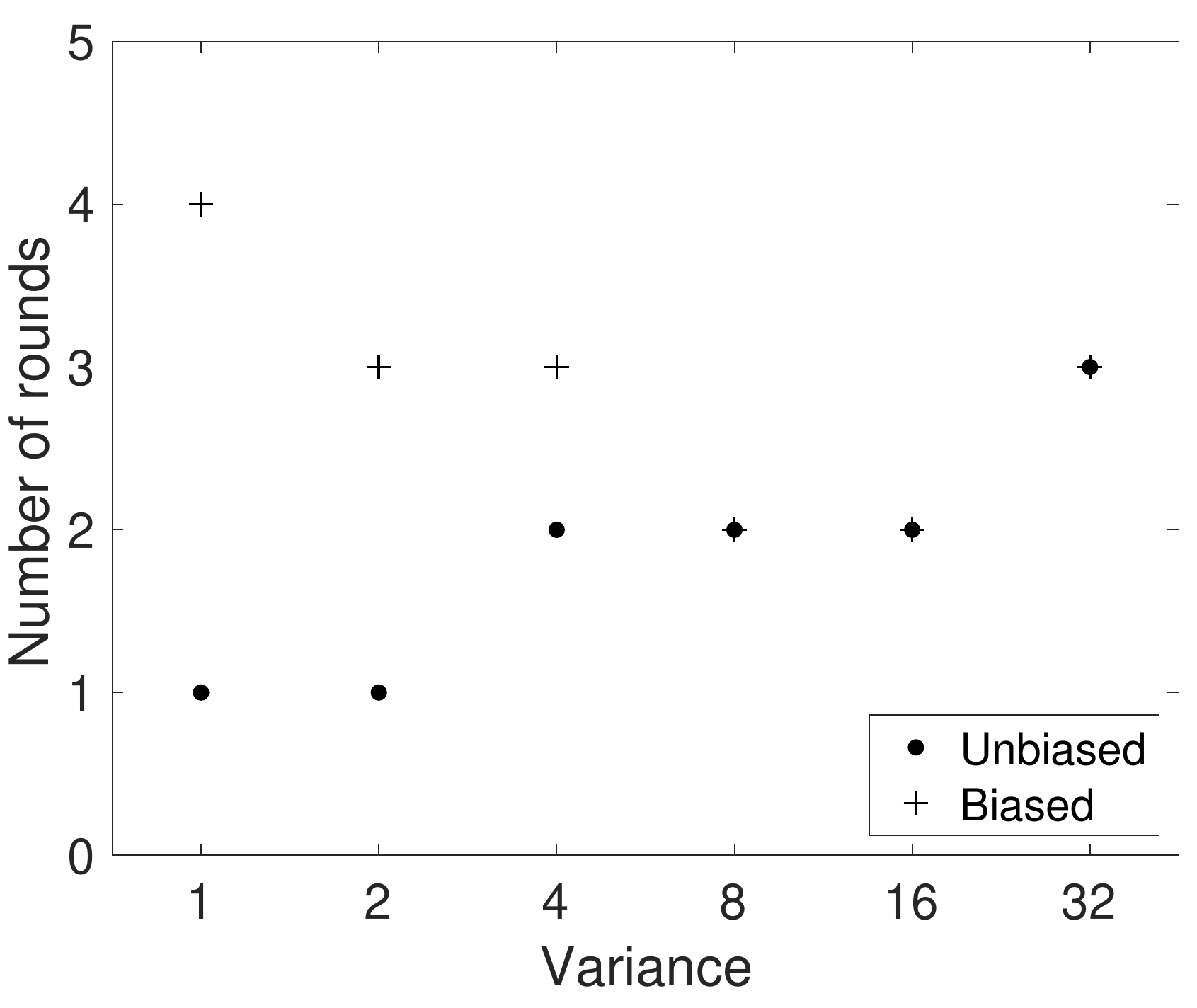}
		\end{center}
		\caption{Auction rounds in LLG.}
		\label{figure:LLG}
	\end{figure}

	Figure~\ref{figure:LLG} plots the number of rounds that our Bayesian
	auction takes to clear the market against the variance
	$\sigma^2_3$ of the prior over the global agent's value. We see that,
	in the unbiased scenario, the number of rounds monotonically increases
	as the variance grows. This can be easily explained since
	increasing the variance only adds noise to the exact prior
	estimate.
	In the biased scenario, we have an optimal range of variances 
	between 8 and 16. If the variance is too low, the auction
	needs many observations to correct the biased prior. If it is too
	high, the low confidence leads to many rounds because the auctioneer
	needs to refine its estimate of the value regardless of the bias.
	
	\paragraph{CATS Experiments}

	For our second set of experiments, we generate instances using the
	Combinatorial Auction Test Suite (CATS), which offers four generator
	distributions: \paths, \regions, \arbitrary, and \scheduling\
	\citep{leyton2000towards}. These are meant to model realistic domains
	such as truck routes, real estate, and pollution rights. We generate
	1000 instances from each distribution, each with 12 items and 10
	single-minded agents.
	
	The instances are generated as follows. First, 100 input files with
	1000 bids each are generated. Each input file is partitioned into a
	``training set'' and ``test set'', each with 500 bids. From the test set,
	10 bids (representing 10 single-minded agents) are sampled uniformly at random. The
	training set is used to fit the prior knowledge of our Bayesian
	auction. Specifically, we fit a linear regression model of bundle
	value according to items contained, using a Gaussian process with a
	linear covariance function, leading to Gaussian prior knowledge. The
	fit is performed using the publicly available GPML Matlab
	code~\citep{williams2006gaussian}.
	
	As a baseline we implemented a standard linear-price auction scheme
	closely related to the combinatorial clock
	auction~\citep{ausubel2014practical}. The scheme is parametrized by a
	positive step size $\tau$. At each round $\ell$, the price of an item
	is incremented by its \emph{excess demand}, scaled by $\tau/\sqrt{\ell}$. The
	excess demand of an item is the number of bidded bundles that
	contain it, minus the number of units of the item offered by the seller at current prices. 
	This can be viewed as a subgradient descent method for computing
	clearing prices, for which a step size proportional to
	$1/\sqrt{\ell}$ yields the optimal worst-case convergence
	rate~\citep[Chap.\ 3]{bertsekas2015convex}.
	
	Both the Bayesian auction and the baseline subgradient auction use
	linear prices, but these may not be expressive enough to support an
	efficient allocation in instances generated by CATS. We therefore set
	a limit of 100 rounds for each auction run, and record reaching this
	limit as a failure to clear the market. 
	
	
	
	\begin{figure}[t!]
		\begin{center}
			\includegraphics[width=0.5\textwidth]{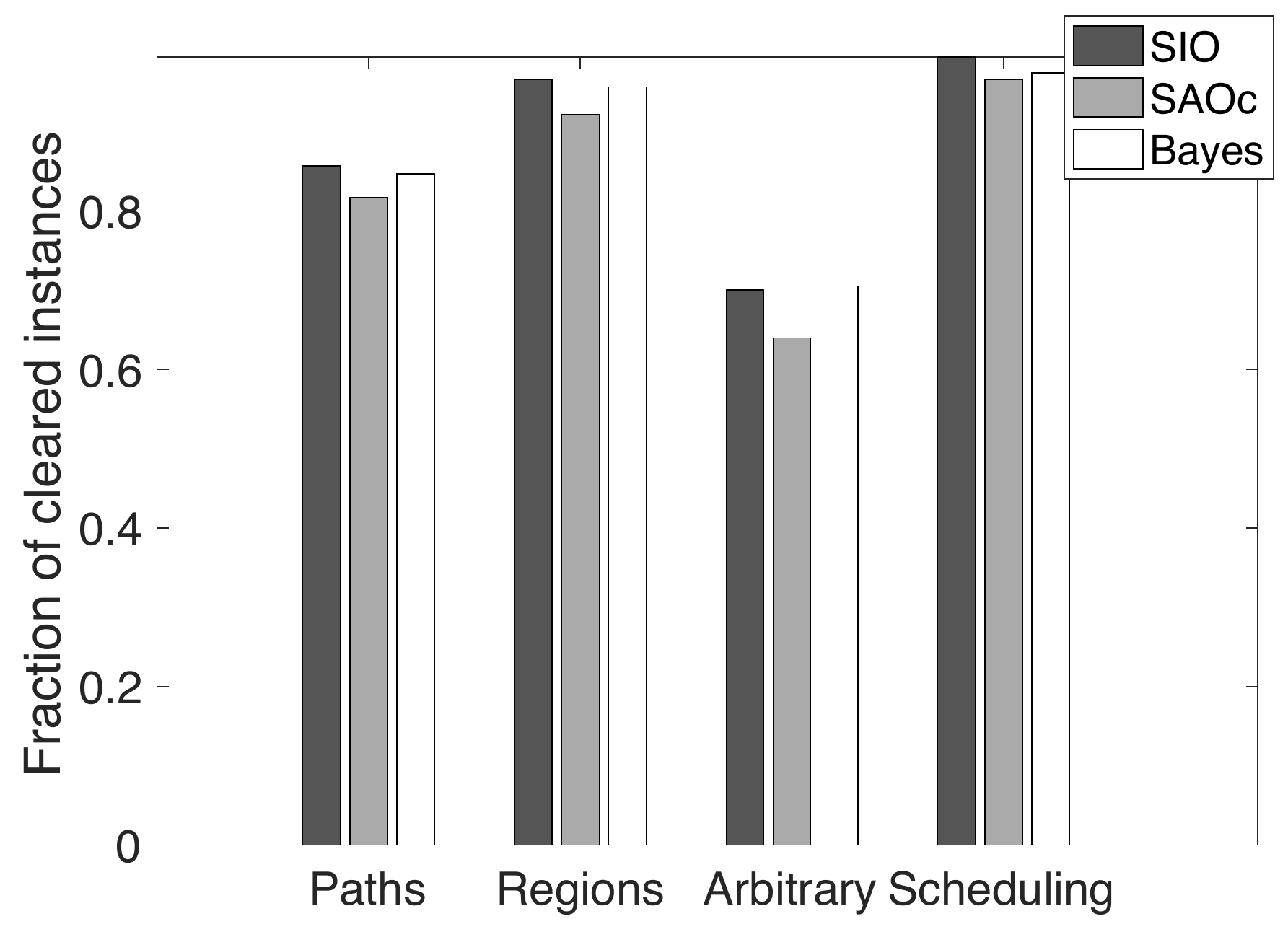}
		\end{center}
		\caption{Cleared instances in CATS.}
		\label{figure:CfrNumOfClearing}
	\end{figure}	
	
	On each instance we run a single Bayesian auction, and 100 subgradient
	auctions with the step size uniformly spanning the interval from zero
	to the maximum agent value. This leads to several baseline results.
	The \emph{standard instance optimized} (SIO) results refer to the
	performance of the baseline when using the optimal step size for each
	instance. The \emph{standard average clearing-optimized} (SAOc)
	results refer to the performance of the baseline under the fixed step
	size that leads to the best clearing performance on average, for each
	valuation domain. Analogously, the \emph{standard average
		round-optimized} (SAOr) results refer to baseline performance under
	the step size leading to lowest average number of rounds.
	{For each instance, the step size that leads to the lowest number of
		rounds naturally leads to the best clearing rate. But the fixed step
		sizes that optimize these two criteria on average may be different.}
	Note that SIO is an extremely competitive baseline, since it is
	optimized for each instance; a priori, we hoped to be competitive
	against it, but did not expect to beat it. The SAOc and SAOr baselines
	reflect more realistic performance that could be achieved in practice.
	
	We first consider clearing performance.
	The results are reported in Figure~\ref{figure:CfrNumOfClearing}. We
	find that the Bayesian auction is competitive with SIO on all four
	domains, and that it always outperforms SAOc. In fact, our auction
	even outperforms SIO on the \arbitrary\ domain. This means that it was
	able to clear some instances that the subgradient auction could not
	clear within 100 rounds at any step size.
	In general, there is good agreement between our Bayesian auction and the 
	baselines on which instances can be cleared or not according 
	to the 100-round criterion. 
	This indicates that failure to clear is typically a property of the instance rather than the algorithm.	
	
\begin{figure}[t!]
		\begin{center}
			\includegraphics[width=0.4\textwidth]{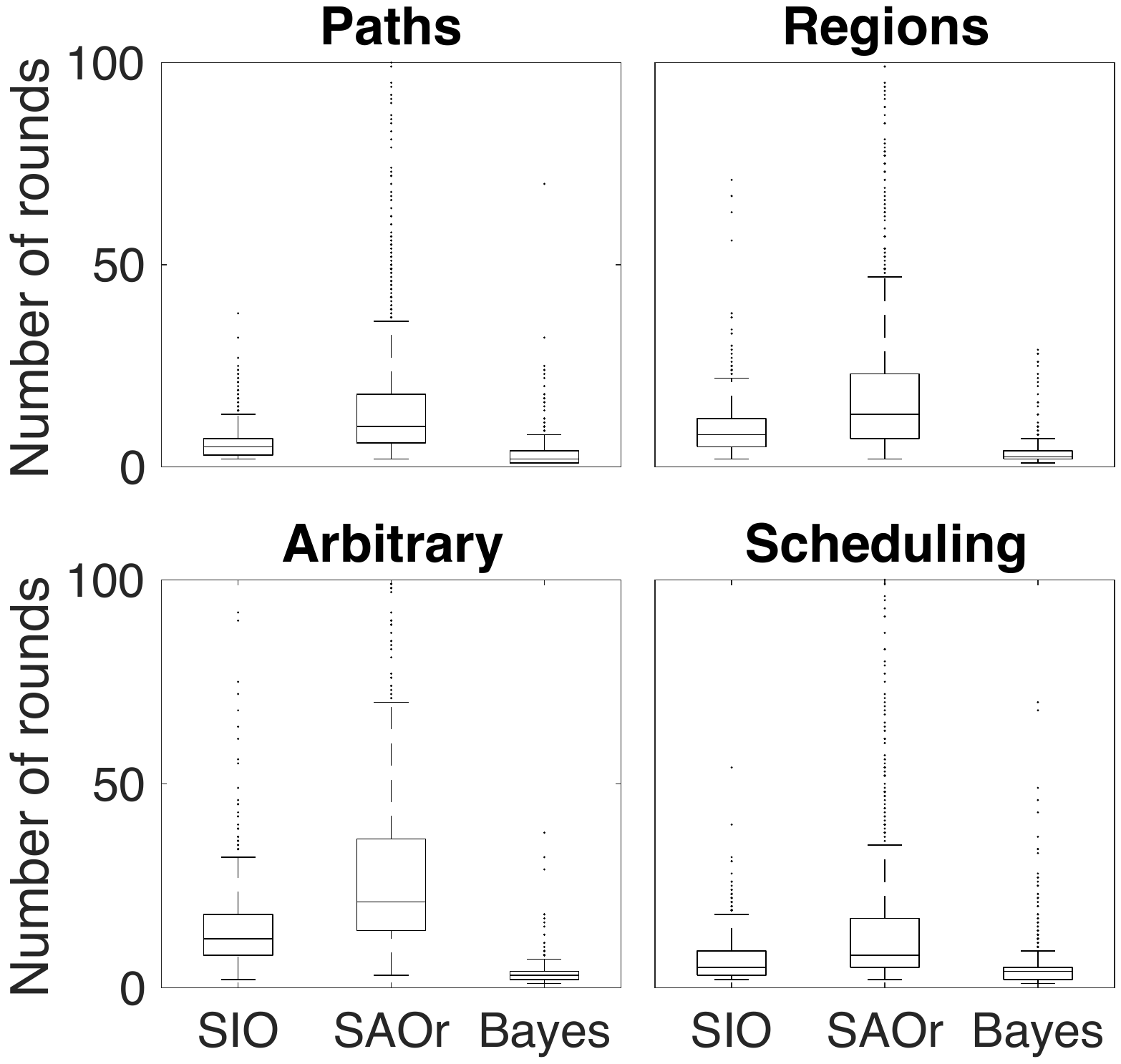}
		\end{center}
		\caption{Auction rounds in CATS.}
		\label{figure:CfrAuctionRounds}
	\end{figure}

	Figure~\ref{figure:CfrAuctionRounds} summarizes the distributions of
	rounds needed to achieve clearing using box plots. To enable fair
	comparisons, for this plot we only considered instances that were
	cleared by all auction types: the Bayesian auction, SAOr, and SIO.
	This yields 770 valid instances for \paths, 910 for \regions, 624 for
	\arbitrary\ and 955 for \scheduling. The mean rounds for the Bayesian
	auction, SAOr, and SIO are always statistically different at the 0.01
	level. We see from the plot that, in terms of the median number of
	rounds, the Bayesian auction clearly outperforms SAOr, but also
	remarkably outperforms SIO. Furthermore, the distribution of rounds
	for the Bayesian auction has a much lower spread than the baselines.
	It is able to clear almost all instances in less than 10 rounds.
	
	%
	
	\section{Conclusion}
	\label{sec:conclusion}

	In this work we developed a Bayesian clearing mechanism for
	combinatorial auctions that allows one to incorporate prior
	information into the auction process in a principled manner.
	Our auction mechanism is based on a joint generative model of
	valuations and prices such that clearing prices are the MAP
	estimate given observed valuations.
	Our empirical evaluation confirmed that our Bayesian mechanism
	performs remarkably well against a conventional clock auction
	scheme, in terms of rounds to convergence. Our auction's
	performance simply relies on reasonable priors for valuations,
	rather than careful tuning of price increments.

	We believe that the Bayesian perspective on auction design
	developed in this paper could be leveraged to improve other
	aspects beyond rounds to convergence. For instance, the
	Bayesian paradigm offers a principled way to select
	hyperparameters~\citep{mackay1992bayesian}; in our context,
	this could be used to choose the right structure of prices
	(linear, nonlinear) to clear the market, a priori. The
	knowledge update component could also form the basis of more
	refined activity rules; for instance, one could reject bids
	that are highly unlikely, given the valuation posterior based
	on previous bids. We intend to pursue these directions in
	future work.

	\bibliographystyle{aaai}
	\bibliography{aaai18}
	
	\newpage
	
	\paragraph{Proof of Proposition~\ref{prop:gen-model}}
	\begin{proof}
		Let $\nu(\bm w, \theta)=\prod_{i} \nu(w_i, \theta) $ and $Q(\bm w) = \prod_{i} Q(w_i)$.\\
		We denote the probability of a restart as 
		\begin{equation}
		r = 1- \int \int \mathrm{d}\bm w \,\mathrm{d}\theta \, \nu(\bm w, \theta) Q(\bm w) Q(\theta),  
		\end{equation}
		where $\mathrm{d}\theta$ is a shorthand for $\prod_{x\in \mathcal X} \mathrm{d}\theta(x)$. (Note that any price function $\theta$ can be seen as a $2^m$-dimensional vector of positive real numbers.) 
		The probability that $(\bm x, \bm w, \theta)$ is drawn after $\ell$ restarts will then be 
		\begin{equation}
		P(\bm x, \bm w, \theta , \ell ) = Q( \bm x \cond \bm w, \theta) Q(\bm w) Q(\theta) \cdot r^{\ell}.
		\end{equation}
		Thus,
		\begin{equation} \begin{array}{lll}
		P(\bm x, \bm w, \theta) &= & \displaystyle \sum_{\ell=0}^{\infty} Q( \bm x \cond \bm w, \theta) Q(\bm w) Q(\theta) \cdot r^{\ell} \\ \\
		&= & \ds \frac{ Q(\bm x \cond \bm w, \theta) Q(\bm w) Q(\theta)}{1-r}.
		\end{array}
		\end{equation}
		Recalling  equation~\eqref{eq:clearing-potential}, we have that 
		\begin{equation} 
			U(\theta; \bx, \bw) \propto Q(\bx \cond \bw, \theta) Q(\theta) .
		\end{equation}		
		We can conclude that 
		\begin{equation}
		P(\bm x, \bm w, \theta) \propto U(\theta; \bx, \bw) Q(\bm w) .
		\end{equation}
	\end{proof}

	\paragraph{Proof of Lemma~\ref{lem:bids-cost-posterior}}	
	\begin{proof}	
	 \begin{eqnarray*}
	  	P(\bb, \bc \cond \bx, \bw, \theta)  & = & \prod_{\ell=1}^k P(\bb^\ell,
	  	\bc^\ell \cond
	  	\bb^{(\ell-1)},
	  	\bc^{(\ell-1)},\bx, \bw, \theta) \\
	  	&\hspace{-1.3cm} = &\hspace{-0.65cm} \prod_{\ell=1}^k
	  	Q(\bb^\ell \cond \bc^\ell, \bw)
	  	P(\bc^\ell \cond
	  	\bb^{(\ell-1)},
	  	\bc^{(\ell-1)}, \bx)\\
	  	&\hspace{-1.3cm} = & \hspace{-0.65cm}
	  	Q(\bb \cond \bc, \bw) \prod_{\ell=1}^k
	  	P(\bc^\ell \cond
	  	\bb^{(\ell-1)},
	  	\bc^{(\ell-1)}, \bx)	  	
	  \end{eqnarray*}
	  The third line follows because bids only depend on bundle costs and
	  values according to~\eqref{eq:bid-behavior}, and costs are independent
	  of values and clearing prices given previous round bids and costs,
	  which fully determine the current round prices.
	\end{proof}

	\paragraph{Proof of Proposition \ref{prop:full-posterior}}	
	\begin{proof}
		We have that
		\begin{eqnarray*}
			P(\bw, \theta \cond \bb, \bc, \bx) & \propto & P(\bb, \bc \cond \bx, \bw, \theta) P(\bx, \bw, \theta) \\
			& \propto & Q(\bb \cond \bc, \bw) P(\bx, \bw, \theta) \\
			& \propto & Q(\bb \cond \bc, \bw) Q(\bx \cond \bw, \theta) Q(\bw) Q(\theta).
		\end{eqnarray*}
		The second line follows from Lemma~\ref{lem:bids-cost-posterior}, and
		the last line from Proposition~\ref{prop:gen-model}.
	\end{proof}	
	
	\paragraph{Price Likelihood Derivation \\ \\}		
	We aim to solve the integral
	\begin{eqnarray*}
		 \int \mathrm{d}w_i\, \hP_i(w_i; m_i, \sigma_i^2) Q(x_i \cond w_i, \theta) = \\
		 & & \hspace{-4.8cm} \frac{1}{2^m}\, \int \mathrm{d}w_i\, \mathcal N(w_i| m_i, \sigma_i^2) \exp{[-(w_i-\theta(x_i))_+]},
	\end{eqnarray*}
	where $\mathcal N(w_i| m_i, \sigma_i^2)$ is the probability density
	function of the normal distribution with mean $m_i$ and variance
	$\sigma_i^2$. We can rewrite the integral as
	\begin{eqnarray*}
		\frac{1}{2^m}\,
		\int_{\theta(x_i)}^{+\infty} \mathrm{d}w_i\, \mathcal N(w_i| m_i, \sigma_i^2) \exp{[-(w_i-\theta(x_i))]}
		+ \\
		&  &  \hspace{-4.5cm} \displaystyle\frac{1}{2^m}\, \int_{-\infty}^{\theta(x_i)} \mathrm{d}w_i\, \mathcal N(w_i| m_i, \sigma_i^2).
	\end{eqnarray*}
	Now, since
	\begin{eqnarray*}
		\mathcal N(w_i| m_i, \sigma_i^2) \exp{[-(w_i-\theta(x_i))]} 
		 =\\ 
		 & & \hspace{-4.8cm}  \mathcal N(w_i| m_i-\sigma_i^2, \sigma_i^2) \exp{[-(m_i-\sigma^2_i/2-\theta(x_i))]},
	\end{eqnarray*}
	and given that $\Phi(-z)=1-\Phi(z)$, we can rewrite our integral as
	\begin{eqnarray*}
		\frac{1}{2^m}\,
		\Phi \left(  \frac{m_i-\theta(x_i)}{\sigma_i} - \sigma_i\right)\, \exp{\left [\theta(x_i) - m_i + \frac{\sigma_i^2}{2}\right ]} + \\ 
		&  &  \hspace{-3.6cm} 
		\displaystyle\frac{1}{2^m}\, \Phi \left(  \frac{\theta(x_i)-m_i}{\sigma_i}
		\right).
	\end{eqnarray*}

\end{document}